\documentclass[12pt]{article}

\newcommand{\be}{\begin{equation}}
\newcommand{\ee}{\end{equation}}
\newcommand{\bi}[1]{\vspace{-3mm} \bibitem{#1}}

\begin{document}

\begin{center}
{\it Physics Letters A 336 (2005) 167-174}

\vskip 3mm

{\Large \bf Continuous Medium Model for Fractal Media}
\vskip 5 mm

{\large \bf Vasily E. Tarasov } \\

\vskip 3mm
{\it Skobeltsyn Institute of Nuclear Physics, \\
Moscow State University, Moscow 119992, Russia}

{E-mail: tarasov@theory.sinp.msu.ru}
\end{center}
\vskip 11 mm

\begin{abstract}
We consider the description of the fractal media that
uses the fractional integrals.
We derive the fractional generalizations of the 
equation that defines the medium mass.
We prove that the fractional integrals can be used to 
describe the media with noninteger mass dimensions.
The fractional equation of continuity is considered. 
\end{abstract}



\vskip 7mm

\section{Introduction}

The application of fractals in physics \cite{It,Main,Zas2}
is far ranging, from the dynamics of fluids in porous media 
to resistivity networks in electronics. 
The cornerstone of fractals is the meaning of dimension, 
specifically the fractal dimension. 
Fractal dimension can be best calculated by box counting 
method which means drawing a box of size R 
and counting the mass inside. 
Fractal models of media are enjoying considerable popularity. 
This is due in part to the relatively 
small number of parameters that can define a 
fractal medium of great complexity and rich structure.

Derivatives and integrals of fractional order have found many
applications in recent studies of scaling phenomena \cite{Zas2,1,2,3,4,MK,Zas}.
It is interesting to use fractional integration to consider 
the properties of the fractional media.
It is interesting to find the connection between
the fractional integrals  and fractals.
The natural questions arise: what could be the physical
meaning of the fractional integration?
This physical meaning can be following:
the fractional integration can be considered as an
integration in some fractional space.
In order to use this interpretation we must define
a fractional space.
The first interpretation of the fractional space
is connected with fractional dimension space.
The fractional dimension interpretation follows from the
formulas for dimensional regularizations.
If we use the well-known formulas for dimensional 
regularizations \cite{Col}, then we get
that the fractional integration can be considered as an 
integration in the fractional dimension space \cite{chaos}
up to the numerical factor
$\Gamma(D/2) /( 2 \pi^{D/2} \Gamma(D))$.
This interpretation was suggested in Ref. \cite{chaos}.

In this Letter we consider the second physical interpretation 
of the coordinate fractional integration.
This interpretation follows from the fractional measure
of space \cite{chaos} that is used in the fractional integrals.
We consider the mass fractional dimension
and the fractional generalizations of the equation 
that defines the mass of the medium.
We prove that the fractional integrals can be used to 
describe fractal media with noninteger mass dimensions.
The fractional integrals can be used not only to calculate 
the mass dimensions of fractal media. 
Fractional integration can be used
to describe the dynamical processes in the fractal media.
Using fractional integrals, we can derive the fractional 
generalization of the dynamical equations.  
The fractional generalization of the Liouville equation 
was suggested in Ref. \cite{chaos}.
Using the Fourier transform, we can introduce fractional dynamical 
equations with coordinate fractional derivatives.

In Section 2, the definition of mass dimension 
and the fractional generalization of the medium mass 
equation is considered. In Section 3, we consider
the properties of the fractal media. 
We define fractality and homogeneity properties of the media.
In Section 4, we discuss the local density of the fractal media.
In Section 5, the fractional generalizations of the equations
are considered. We prove that these equations can be used
to describe fractal media.  
In section 6, we consider the fractional equation of continuity. 
Finally, a short conclusion is given in Section 7.

\section{Mass Fractal Dimension}

Equations that define the fractal dimensions have 
the passage to the limit. This passage makes difficult 
the practical application to the real fractal media.
The other dimensions, which can be easy calculated from
the experimental data, are used in empirical investigations.
For example, the mass fractional dimension \cite{Mand,Schr} 
can be easy measured.

The properties of the fractal media like mass obeys a power law relation
\be \label{MR} M(R) =kR^{D_m} , \ee
where $M$ is the mass of fractal medium, $R$ is a box size (or a sphere radius),
and $D_m$ is a mass fractal dimension. 
Amount of mass of a medium inside a box of size $R$
has a power law relation (\ref{MR}).

Fractal dimension can be calculated by box counting method
which means drawing a box of size R 
and counting the mass inside. 
To calculate the mass fractal dimension, take the logarithm (\ref{MR})
both sides
\[ ln (M)=D_m \ ln(R)+ln k . \]
Log-log plot of $M$ and $R$ gives us the slope $D_m$, 
the fractal dimension. 
When we graph ln(M) as a function of ln(R), we get a value 
of about $D_m$ which is the fractal dimension of fractal media.

The power law relation (\ref{MR}) can be naturally 
derived by using the fractional integral.

In order to describe the fractal media, we suggest to use the 
space with fractional measure. 
In this case, we can use the constant density distribution
for homogeneous fractal media.  
If we use $\rho(x)=\rho_0=const$, then we get (\ref{MR}). 
In the next sections, we prove that the mass fractal dimension 
is connected with the order of fractional integrals. 
Therefore the fractional integrals can be used to describe fractal
media with non-integer mass dimensions.

\section{Fractal Media}

Let us consider the region $W_A$ in 3-dimensional 
Euclidean space $E^3$, where $A$ is the midpoint of this region.
The volume of the region $W_A$ is denoted by $V(W_A)$.
If the region $W_A$ is a ball with the radius $R_A$,
then the midpoint $A$ is a center of the ball, and
the volume $V(W_A)=(4/3)\pi R^3_A$ .

The mass of the region $W_A$ in the fractal media is denoted 
by $M(W_A)$. Let $\bar{\rho}(W_A)$ is an average mass density
of the region $W_A$ of the fractal medium. 
This density is defined by the equation
\[ \bar{\rho}(W_A)=M(W_A)/V(W_A) . \]

In the general case, the fractal media cannot be considered as continuous media.
There are points and domains that do not filled by the medium particles.
These domains can be called the porous. 
The fractal media can be considered as continuous media
for the scales much more than mean value of the pore size $R_p$.

The fractal media can be characterized by the following property: \\

{\it If the volume $V(W_A)$ of the region $W_A$ increase,
then the average mass density $\bar{\rho}(W_A)$ is decrease.} \\

This property is satisfied for all points $A$ and all regions $W_A$
if the volume of region is much more the
average value of porous volume ($V(W_A)>>V_p$).
For the ball region of the fractal media, 
this property can be described by the power law 
\[ {\bar \rho}(W_A)=\rho_0 (R/R_p)^{D-3} , \]
where $\rho_0$ is a constant value;
$R$ is the radius of the ball $W_A$.
Here $R>>R_p$, where $R_p$ is a mean radius of the porous sphere.

The fractality of medium means that the mass of this medium 
in any region of Euclidian space $E^3$ increase more slowly 
than the volume of this region.
For the ball region $W$, we have
\[ M(R) \sim R^{D}, \ \ \ D<3 . \]

The characteristic property of fractal media can be 
considered in the following forms: \\

{\it For all regions $W_A$ and $W_B$ in the fractal media such that 
$W_A\subset W_B$ \\
and  $V(W_A)<V(W_B)$, we have that the corresponding 
average mass densities satisfy the inequality 
${\bar \rho}(W_A)>{\bar \rho}(W_B)$, i.e., }
\be \label{F} W_A\subset W_B, \ \ V(W_A)<V(W_B) \Rightarrow
{\bar \rho}(W_A)>{\bar \rho}(W_B) . \ee

We would like to consider the fractal medium that have some homogeneous
property. Now we shall give the definition of homogeneous fractal media.

The fractal media are called homogeneous if the following property
is satisfied: \\

{\it For all regions $W_A$ and $W_B$ of the homogeneous fractal media
such that the volumes are equal $V(W_A)=V(W_B)$, 
we have average densities of these regions are equal too \\
${\bar \rho}(W_A)={\bar \rho}(W_B)$, i.e., } 
\be \label{F0}
V(W_A)=V(W_B) \ \ \Rightarrow \ \ {\bar \rho}(W_A)={\bar \rho}(W_B) 
\ee

Fractal medium is called a homogeneous fractal medium if
the average density value of the region 
does not depends on the translation of this region. 

Note that the wide class of the fractal media satisfies 
the homogeneous property.
In many cases, we can consider the porous media \cite{Por1,Por2}, 
polymers \cite{P}, colloid agregates \cite{CA}, and 
aerogels \cite{aero} as homogeneous fractal media.
The dendrites cannot be considered as a homogeneous fractal medium. 

We can generalize the fractal property (\ref{F})
for the homogeneous fractal media.
This generalization is connected with consideration of
two different points and two different regions of the media.
For the homogeneous fractal media we can remove the
restriction $W_A \subset W_B$ in Eq. (\ref{F}).
These regions can satisfy the condition $W_A \cap W_B = \emptyset$.
Therefore we have the following property: \\

\noindent
{\it For all regions $W_A$ and $W_B$ of the homogeneous fractal 
medium in the Euclidean space $E^3$ that
satisfies the inequality $V(W_A)<V(W_B)$, 
we have that the average densities are connected 
by the inequality ${\bar \rho}(W_A) >{\bar \rho}(W_B)$.} \\

As the result, we have the following properties for all 
regions $W_A$ and $W_B$ in the Euclidean space $E^3$ with homogeneous 
fractal media: \\

\noindent
(1) {\it Fractality}:
if $V(W_A)<V(W_B)$, then ${\bar \rho}(W_A) >{\bar \rho}(W_B)$. \\

\noindent
(2) {\it Homogeneity}: 
if $V(W_A)=V(W_B)$, then ${\bar \rho}(W_A) ={\bar \rho}(W_B)$ and 
$M(W_A)=M(W_B)$. \\

\section{Local Density of Fractal Media}

In the general case, the fractal media cannot be considered as continuous media.
There are points and domains that do not filled of the medium particles.
These domains can be called the porous. 
The fractal media can be considered as continuous media
for the scales much more than mean value of the pore size $R_p$.
In this case, we can use the integration and differentiation.

In many cases, the fractal media are described by the equations
with the integer integration. However, this description
can be incorrect for fractal media. 
For example, the mass of the medium is derived by the 
following equation
\be \label{MW} M(W)=\int_W \rho({\bf r}) d^3 {\bf r} . \ee
If we describe local density $\rho({\bf r})$ as the single-valued 
function, then we cannot use Eq. (\ref{MW}) with the integer integral
for the homogeneous fractal media.
%%%If we would like to use this equation with the 
%%%integer integral, then we cannot define the local mass 
%%%density $\rho({\bf r})$ for the homogeneous fractal media.

In order to prove this statement, we consider the 
points $A$, $B$, and $C$, such that $R_A=|AC| > R_B=|BC|$. 
Obviously,  we require that $R_A >> R_p$ and $R_B >> R_p$. 
Let us describe the mass of the media using the integer integrals.
The mass of the regions $W_A$ and $W_B$ of 
the homogeneous fractal media are defined by the equations
\be \label{MWAB} M(W_A)=\int_{W_A} \rho_A({\bf r}) d^3 {\bf r} , \quad 
M(W_B)=\int_{W_B} \rho_B({\bf r}) d^3 {\bf r} . \ee
We use the index $A$ and $B$ in the local densities, since we
would like to prove that $\rho_A({\bf r}_C)\not=\rho_B({\bf r}_C)$.
Obviously, that the local densities $\rho_A$ and $\rho_B$ of the continuous 
medium satisfy the condition $\rho_A({\bf r}_C)=\rho_B({\bf r}_C)$. 

In order to satisfy the fractal property of the medium and Eqs. (\ref{MWAB}),
we must use the following local densities:
\be \label{rAB}
\rho_A({\bf r})=\rho_0 \Bigl( |{\bf r}-{\bf r}_A| / R_p \Bigr)^{D-3},
\quad 
\rho_B({\bf r})=\rho_0 \Bigl( |{\bf r}-{\bf r}_B| / R_p \Bigr)^{D-3} .
\ee

Let us consider the point $C$ such that $|AC| > |BC|$.
Using Eq. (\ref{rAB}), we have the local density in
this point in the form
\[ \rho_A({\bf r}_C)=\rho_0 \Bigl( |AC| / R_p \Bigr)^{D-3},
\quad  \rho_B({\bf r}_C)=\rho_0 \Bigl( |BC| / R_p \Bigr)^{D-3}. \]
Since $|AC| > |BC|$, it follows that $\rho_A({\bf r}_C)<\rho_B({\bf r}_C)$. 
Obviously, that we have two different values for one point $C$ of 
fractal medium.
Therefore we cannot use Eq. (\ref{MW}) with the integer integrals
for the homogeneous fractal media.
We can use Eq. (\ref{MW}) with the local density
$\rho({\bf r}) \sim |{\bf r}|^{D-3}$ 
only for the non-homogeneous distribution of the medium mass 
that has the single out point. This point is defined by ${\bf r}=0$. 

As the result, we have that the fractality and homogeneity properties
of the medium cannot be 
satisfied simultaneously by Eq. (\ref{MW}) with the integer integrals.

In order to satisfy these properties, we must use the 
generalization of Eq. (\ref{MW}) such that fractality 
and homogeneity properties can be realized in the form: \\

\noindent
(1) {\bf Fractality}:
the mass of the region $W$ of fractal medium obeys a power law relation
\be \label{MR2} M(W)=M_0 \Bigl( \frac{(V(W_A))^{1/3}}{R_p} \Bigr)^D , \ee
where $D<3$. For the ball region $W_A$, we have
\[ M(R)=M_0(R/R_p)^D , \]
where $R=|{\bf r}-{\bf r}_A|$. \\

\noindent
(2) {\bf Homogeneity}:
the local density of homogeneous fractal medium
is translation invariant value that have the form
\be \label{const} \rho({\bf r})=\rho_0=const . \ee

\vskip 3mm

We can realize these requirements by the equation 
\[ M(W)=\int_W \rho({\bf r}) d\mu_D({\bf r}) , \]
where $\mu_D$ is a new measure of the space $E^3$. 
Using Eq. (\ref{const}), we have
\[ M(W)=\rho_0\int_W d\mu_D({\bf r})=\rho_0 V_D(W) , \]
where $V_D(W)$ is the volume of the region $W$.
Therefore we get that the homogeneous fractal media 
are described by the measure space $(E,\mu)$ such that
\[ V_D(W)=\frac{\rho_0}{M_0} \Bigl(\frac{V(W)}{R_p}\Bigr)^{D/3} , \]
where $V(W)$ is the usual volume of the region $W$,
i.e., we have $V_D(W)\sim (V(W))^{D/3}$. 
If we consider the ball region, we get
\[ V_D(W)=V_0 (R/R_p)^{D} , \]
where $D<3$, and $V_0=(4 \pi /3)^{D/3}(\rho_0/M_0)$.
In the next section, we prove that the natural generalization
of Eq. (\ref{MW}) uses the fractional integration.

\section{Fractional Equations}

Let us prove that the fractional integration allows us to
generalize Eq. (\ref{MW}) such that the 
fractality and homogeneity conditions are realized in the form
\[ M(W) \sim (V(W))^{D/3} , \quad V_D(W) \sim (V(W))^{D/3}, \quad 
\rho({\bf r})=const . \]
%%%are satisfied.

Let us define the fractional integral
in Euclidean space $E^3$ in the Riesz form
%%%The Rietz integration in Euclidean space $E^3$ is defined 
\cite{SKM} by the equation
\be \label{ID} (I^{D}\rho)({\bf r}_0)=\gamma^{-1}_3(D)
\int_W \frac{\rho({\bf r}) d^3 {\bf r}}{|{\bf r}-{\bf r}_0|^{3-D}} , \ee
where ${\bf r}_0 \in W$, and we use
\[ |{\bf r}-{\bf r}_0|=\sqrt{(x-x_0)^2+(y-y_0)^2+(z-z_0)^2} ,\quad
\gamma_3(D)= \frac{2^D \pi^{3/2} \Gamma(D/2)}{\Gamma(1/2)} .\]
Using ${\bf R}={\bf r}-{\bf r}_0$, we can be rewritten Eq. (\ref{ID})
in an equivalent form
\[ (I^{D}\rho)({\bf r}_0)=\gamma^{-1}_3(D)
\int_W \rho({\bf R}+{\bf r}_0) |{\bf R}|^{D-3} d^3 {\bf R} . \]

Note that $\gamma_3(3-0)\not=1$, where
\[ \gamma_3(3-0)=\lim_{D\rightarrow 3-0} \gamma_3(D) =
 \frac{2^3 \pi^{3/2} \Gamma(3/2)}{\Gamma(1/2)} . \]
Using notations (\ref{ID}), 
we can write Eq. (\ref{MW}) in the form
\[ M(W)=\gamma_3(3-0)(I^{3}\rho)({\bf r}_0) . \]
The fractional generalization has the form
\[ M(W)=\gamma_3(3-0)(I^{D}\rho)({\bf r}_0) . \]
We use the factor $\gamma_3(3-0)$ to derive Eq. (\ref{MW}) 
in the limit $D \rightarrow 3-0$. 
The fractional equation for $M(W)$ can be written in the equivalent form
\be \label{MWD}  M(W)= \frac{2^{3-D} \Gamma(3/2)}{\Gamma(D/2)}
\int_W \rho({\bf R}+{\bf r}_0) |{\bf R}|^{D-3} d^3 {\bf R} , \ee
where we use
\[ \gamma_3(3-0) \gamma^{-1}_3(D)=\frac{2^{3-D} \Gamma(3/2)}{\Gamma(D/2)} . \]
Here and late we use the initial points in the integrals are set to zero.

If we have $\rho({\bf r})=\rho_0=const$ and the ball region $W$, then
\[ M(W)= \rho_0 \frac{2^{3-D} \Gamma(3/2)}{\Gamma(D/2)} 
\int_W |{\bf R}|^{D-3} d^3 {\bf R} . \]
Using the spherical coordinates, we get
\[ M(W)= \frac{\pi 2^{5-D} \Gamma(3/2)}{\Gamma(D/2)} \rho_0 \int_W R^{D-1} d R= 
\frac{2^{5-D} \pi \Gamma(3/2)}{D \Gamma(D/2)} \rho_0 R^{D} , \]
where $R=|{\bf R}|$. 
As the result, we have $M(W)\sim R^D$, i.e., we derive Eq. (\ref{MR2})
up to the numerical factor.

We can suppose that Eq. (\ref{MWD}) can be used for non-homogeneous
fractal media. In this case, the distribution
depends on $R=|{\bf R}|$, and the angels.
%%%, and the single out point,
%%%which is defined by the vector ${\bf r}_0$.

Let us consider the spherical-homogeneous fractal media
that have the single out point.
These media are defined by the following properties: \\

\noindent
(1) {\it For all regions $W_A$ and $W_B$ in the Euclidian space $E^3$
such that $|OA|=|OB|=R$ and $R$ much more than the size of the regions
($R^3_p << V(W_A)=V(W_B) << R^3$), we have the relation
${\bar \rho}(W_A)={\bar \rho}(W_B)$. } \\

\noindent
(2) {\it For all regions $W_A$ such that
the size of the region is much smaller than distance $|OA|=R$ 
($R^3_p << V(W_A)<< R^3$), we have that the density ${\bar \rho}(W_A)$ 
depends on the distance $R=|OA|$.} \\

For example, we can consider the power law in the form
\be \label{OAb} {\bar \rho}(W_A) \sim |OA|^{\beta} . \ee
The example of the spherical-homogeneous fractal media
we can be realized in nature as dendrites.

If we consider the non-homogeneous fractal media, then the local density 
$\rho=\rho({\bf R})$ have the single out point.
%%%which is defined by the vector ${\bf r}_0$.
For the spherical symmetric case,
we have $\rho({\bf R})=\rho(|{\bf R}|)$
and the equation can be represented in the form
\[ M(W)= \frac{\pi 2^{5-D} \Gamma(3/2)}{\Gamma(D/2)}
\int^R_0  \rho(R) R^{D-1} d R . \]
%%%where $R=|{\bf R}|$. 
For the simple case (\ref{OAb}), we have
%%%\be \label{MWD2}  M(W)=\gamma_3(3-0) \gamma^{-1}_3(D)
%%%\int_W \rho({\bf R}) |{\bf R}|^{D-3} d^3 {\bf R} . \ee
the distribution in the form
\[\rho({\bf R})=\rho(|{\bf R}|)=c_0 |{\bf R}|^{\beta} . \]
Therefore the mass of fractal medium in the region $W$ is
\[  M(W)=c_0 \frac{2^{3-D} \Gamma(3/2)}{\Gamma(D/2)}
\int_W |{\bf R}|^{D+\beta-3} d^3 {\bf R} . \]
Using the spherical coordinates, we have
\[ M(W)= c_0 \frac{\pi 2^{5-D} \Gamma(3/2)}{\Gamma(D/2)} \int^R_0 R^{D+\beta-3} d R = 
\frac{\pi 2^{5-D}  \Gamma(3/2)}{(D+\beta) \Gamma(D/2)} R^{D+\beta} . \]
In this case, the mass-dimension is equal to $D_m=D+\beta$. 

To calculate $\beta$, we can use the definition of the spherical-homogeneous 
fractal media and the generalization of the
box counting method. This generalization must use the box regions $W$
that satisfy the conditions $V(W) <<R$ from the definition 
of spherical-homogeneous fractal media.

\section{Fractional Equation of Continuity}

The fractional integrals can be used not only to calculate 
the mass dimensions of fractal media. 
Fractional integration can be used
to describe the dynamical processes in the fractal media.
Using fractional integrals, we can derive the fractional generalization 
of dynamical equations \cite{chaos}.  
Let us derive the fractional analog of the equation of continuity
for the fractional media.

Let us consider a domain $W_{0}$ for the time $t=0$.
In the Hamilton picture  we have
\[ \int_{W_{t}} \rho({\bf R}_{t},t) d\mu_{D} ({\bf R}_{t})=
\int_{W_{0}} \rho({\bf R}_{0},0) d\mu_{D} ({\bf R}_{0}), \]
where we use the following notation
\[ d\mu_{D} ({\bf R})= \frac{2^{3-D} \Gamma(3/2)}{\Gamma(D/2)}
|{\bf R}|^{D-3} d^3 {\bf R}.  \]
Using the replacement of variables ${\bf R}_{t}={\bf R}_{t}({\bf R}_{0})$,
where ${\bf R}_{0}$ is a Lagrangian variable, we get
\[ \int_{W_{0}}  \rho({\bf R}_{t},t) |{\bf R}_t|^{D-3}
\frac{\partial {\bf R}_{t}}{\partial {\bf R}_{0}} d{\bf R}_{0}=
\int_{W_{0}}  \rho({\bf R}_{0},0) |{\bf R}_0|^{D-3} d{\bf R}_{0}, \]
where $\partial {\bf R}_t/ \partial {\bf R}_0$ is Jacobian. 
Since $W_{0}$ is an arbitrary domain we have
\[ \rho ({\bf R}_{t},t) d\mu_{D} ({\bf R}_{t})=
\rho ({\bf R}_{0},0) d\mu_{D} ({\bf R}_{0}), \]
or the equivalent form
\[ \rho({\bf R}_{t},t) |{\bf R}_t|^{D-3}
\frac{\partial {\bf R}_{t}}{\partial {\bf R}_{0}} =
\rho({\bf R}_{0},0) |{\bf R}_0|^{D-3}_{0} . \]
Differentiating this equation in time $t$, we obtain
\[ \frac{d  \rho({\bf R}_{t},t)}{dt} |{\bf R}_t|^{D-3}
\frac{\partial {\bf R}_{t}}{\partial {\bf R}_{0}}
+  \rho({\bf R}_{t},t) \frac{d}{dt} \Bigl(|{\bf R}_t|^{D-3}
\frac{\partial {\bf R}_{t}}{\partial {\bf R}_{0}}\Bigr)=0 , \]
\be \label{Liu} \frac{d  \rho({\bf R}_{t},t)}{dt}
+ \Omega_{D}({\bf R}_{t},t)  \rho({\bf R}_{t},t)=0 , \ee
where $d/dt$ is a total time derivative
\[ \frac{d}{dt}=\frac{\partial}{\partial t}+\frac{d {\bf R}_t}{dt}
\frac{\partial}{\partial {\bf R}_{t}} . \]
The function
\[ \Omega_{D}({\bf R}_{t},t)=
\frac{d}{dt} ln \Bigl(|{\bf R}_t|^{D-3}
\frac{\partial {\bf R}_{t}}{\partial {\bf R}_{0}}\Bigr) \]
describes the velocity of volume change. 
%%%Indeed, we can derive \cite{chaos} the relation
%%%\[ \frac{dV_D}{dt}=\int_W \Omega_D d\mu_D . \]
Eq. (\ref{Liu}) is a fractional Liouville equation
in the Hamilton picture. 
Using the equation 
\[ \frac{d{\bf R}_{t}}{dt}={\bf V}_t({\bf R},t), \]
we get the function 
\[ \Omega_{D}({\bf R}_{t},t)=
\frac{d}{dt} \Bigl( ln \ |{\bf R}_t|^{D-3} +
ln \ \frac{\partial {\bf R}_{t}}{\partial {\bf R}_{0}}\Bigr)=
(D-3) \frac{1}{|{\bf R}_{t}|} \frac{d|{\bf R}_{t}|}{dt} +
\frac{\partial}{\partial {\bf R}_{t}} \frac{d{\bf R}_{t}}{dt} . \]
As the result we have
\[ \Omega_{D}({\bf R},t)=\frac{(D-3) ({\bf R},{\bf V}_t)}{|{\bf R}|^2} +
\frac{\partial {\bf V}_t}{\partial {\bf R}}, \]
where we use 
$\partial |{\bf R}_t| / \partial {\bf R}_t={\bf R}_t/|{\bf R}_t|$.
%%%Here \[ ({\bf R},{\bf V})=\sum^3_{k=1} R_k V_k . \]

In the general case ($D \not=3$), the function
$\Omega_{D}$ is not equal to zero ($\Omega_{D} \not=0$)
for solenoidal velocity fields 
($div {\bf V}=0$).
If $D=3$, we have $\Omega_{D} \not=0$ only
for non-solenoidal velocity field. 
Therefore the flow in the fractional medium is similar 
to the non-solenoidal flow. 

As the result, we have the following equation of continuity for
fractal media:
\[ \frac{\partial \rho({\bf R}_{t},t)}{\partial t}+{\bf V}_t
\frac{\partial \rho({\bf R}_{t},t)}{\partial {\bf R}_{t}} +
\frac{(D-3) ({\bf R_t},{\bf V}_t)}{|{\bf R}_{t}|^2}\rho({\bf R}_{t},t)+
\rho({\bf R}_{t},t)\frac{\partial {\bf V}_{t}}{\partial {\bf R}_{t}}=0 . \]
For the homogemeous fractal media, we have $\rho=const$ and 
the equation of continuity leads us to the equation
\[ \frac{(D-3) ({\bf R_t},{\bf V}_{t})}{|{\bf R}_{t}|^2}+
\frac{\partial {\bf V}_{t}}{\partial {\bf R}_{t}}=0 . \]
Therefore, we get the non-solenoidal flow of the velocity 
$div {\bf V}=\partial {\bf V}/ \partial {\bf R} \not=0$.

\section{Conclusion}

The application of fractals in physics is far ranging, from the 
dynamics of fluids in porous media to resistivity 
networks in electronics. 
The cornerstone of fractals is the meaning of dimension, 
specifically the fractal dimension. 
Fractal dimension can be best calculated by box counting 
method which means drawing a box of size R 
and counting the mass inside. 
When we graph ln(M) as a function of ln(R) we get a value 
of about $D_m$ which is the fractal dimension of fractal media. 
The mass fractal dimensions of the media 
can be easy measured by experiments. 
The experimental determination of the mass fractal dimension 
can be realized by the usual  box counting methods.
The mass dimensions were measured for porous materials
\cite{Por1,Por2}, polymers \cite{P}, and colloid agregates \cite{CA}.
For the porous materials the fractal dimension is proportional to
porosity of the medium. 
Fractal models of porous media are enjoying considerable
popularity \cite{AT,PMRM,BD,PBR,RP,RC}. 
This is due in part to the relatively 
small number of parameters that can define a fractal porous medium 
of great complexity and rich structure.

We suppose that the concept of fractional integration
provides an alternative approach to describe the fractal media. 
In this Letter we prove that mass fractal media can be described
by using the fractional integrals. 
We consider the fractional generalizations of 
the equation that defines the medium mass.
This fractional generalization allows us to realize the 
consistent description of the fractal mass dimension of the media. 

The fractional integrals can be used in order to describe 
the dynamical processes in the fractal media. The physics of 
fractal media can be successful described by the fractional integration. 
Fractional integration approach is potentially more useful 
for physics of fractal media than traditional methods that 
use the integer integration.

Note that the Liouville equation is a cornerstone of 
the statistical mechanics \cite{Is,RL,F}.  
The fractional generalization of the Liouville equation 
was suggested in Ref. \cite{chaos}. 
This fractional Liouville equation was derived from 
the fractional generalization of the normalization 
condition \cite{chaos} that uses fractional integrals. 
The fractional generalization of Liouville equation allows us 
to derive the fractional analogs of 
Bogoliubov equations \cite{PRE05}, Fokker-Planck equation, 
Enskog transport equation, and hydrodynamic equations. 
Using the Fourier transform, we can get the fractional dynamical 
equations with coordinate fractional derivatives.
Using the methods suggested in Refs. \cite{Tarpla1,Tarmsu,Tarsam},
we can realize the Weyl quantization of the suggested 
fractional equations.

%%%%%%%%%%%%%%%%%%%%%%%%%%%%%%%%%%%%%%%%%%%%%%%%%%%%%%%%%%%

%%%%%%%%%%%%%%%%%%%%%%%%%%%%%%%%%%%%%%%%%%%%%%%%%%%%%%%%%%%%%%%%%%%%%%%%%%

\end{document}